\documentclass[conference]{IEEEtran}
\usepackage{lipsum}
\usepackage{amsthm}
\usepackage{framed} 

\newtheoremstyle{mystyle}
  {}
  {}
  {\itshape}
  {}
  {\bfseries}
  { }
  {.5em}
  {}

\theoremstyle{mystyle}

\usepackage{ifpdf}

\usepackage{cite}

\ifCLASSINFOpdf
\else
\fi
\usepackage{fixltx2e}

\usepackage{stfloats}
\usepackage{amsmath,dsfont,bbm,epsfig,amssymb,amsfonts,amstext,verbatim,amsopn,cite,subfigure,multirow,multicol,lipsum,xfrac}
\usepackage{mathtools}
\usepackage{perpage}
\usepackage{balance}
\usepackage{url}
\usepackage{amsfonts}
\usepackage{epsfig}
\usepackage[font={small}]{caption}
\usepackage{slashbox}
\usepackage{setspace}
\usepackage{stmaryrd}
\usepackage{psfrag}	
\usepackage{multirow}
\usepackage{slashbox}
\usepackage[process=auto]{pstool}
\usepackage{etoolbox}
\usepackage{algorithmicx}
\usepackage[Algorithm,ruled]{algorithm}
\usepackage{algpseudocode}
\usepackage{pifont}
\usepackage[utf8]{inputenc}
\usepackage[T1]{fontenc}  
\usepackage[nolist]{acronym}
\MakePerPage{footnote}
\usepackage{paralist}
\usepackage{enumitem}
\usepackage{bbm}
\usepackage{tikz}
\usetikzlibrary{shapes,arrows}

\newcommand{\maX}{\mathbbmss{X}}

\newcommand{\mae}{\mathcal{E}}
\newcommand{\maz}{\mathcal{Z}}

\newcommand{\man}{\mathcal{N}}

\newcommand{\mg}{\mathcal{G}}
\newcommand{\mai}{\mathcal{I}}
\newcommand{\maw}{\mathbbmss{W}}

\newcommand{\rate}{\mathrm{r}}
\newcommand{\Mrm}{\mathsf{M}}

\newcommand{\moment}{\mathsf{M}}
\newcommand{\lams}{\lambda^{\mathsf{s}}}

\newcommand{\real}{\mathbbmss{R}}

\newcommand{\bgg}{\mathbf{g}}
\newcommand{\prm}{\mathrm{p}}

\newcommand{\bx}{{\boldsymbol{x}}}
\newcommand{\hx}{{\hat{x}}}
\newcommand{\hv}{{\hat{v}}}

\newcommand{\bhx}{{\boldsymbol{\hat{x}}}}

\newcommand{\bxx}{\mathbf{x}}
\newcommand{\bvv}{\mathbf{v}}

\newcommand{\bz}{{\boldsymbol{z}}}

\newcommand{\bv}{{\boldsymbol{v}}}

\newcommand{\dif}{\mathrm{d}}

\newcommand{\trp}{\mathsf{T}}

\newcommand{\ba}{\mathbf{a}}
\newcommand{\by}{{\boldsymbol{y}}}

\newcommand{\mA}{\mathbf{A}}

\newcommand{\mI}{\mathbf{I}}
\newcommand{\mone}{\mathbf{1}}
\newcommand{\mJ}{\mathbf{J}}
\newcommand{\mG}{\mathbf{G}}
\newcommand{\mQ}{\mathbf{Q}}

\newcommand{\mU}{\mathbf{U}}

\newcommand{\mD}{\mathbf{D}}
\newcommand{\mM}{\mathbf{M}}

\newcommand{\mT}{\mathbf{T}}
\newcommand{\md}{\mathrm{D}}
\newcommand{\E}{\mathsf{E}\hspace{.5mm}}

\newcommand{\norm}[1]{\lVert #1 \rVert}
\newcommand{\abs}[1]{\lvert #1 \rvert}

\newcommand{\Tr}{\mathrm{Tr} }

\newtheorem{proposition}{Proposition}

\newcounter{bar}

\begin{document}
\label{list:acronyms}
\begin{acronym}
\acro{iid}[i.i.d.]{independent and identically distributed}
\acro{pmf}[PMF]{Probability Mass Function}
\acro{cdf}[CDF]{Cumulative Distribution Function}
\acro{pdf}[PDF]{Probability Density Function}
\acro{rs}[RS]{Replica Symmetry}
\acro{1rsb}[1RSB]{One-Step Replica Symmetry Breaking}
\acro{brsb}[$b$RSB]{$b$-Steps Replica Symmetry Breaking}
\acro{rsb}[RSB]{Replica Symmetry Breaking}
\acro{mse}[MSE]{Mean Square Error}
\acro{mmse}[MMSE]{Minimum Mean Square Error}
\acro{map}[MAP]{Maximum-A-Posteriori}
\acro{awgn}[AWGN]{Additive White Gaussian Noise}
\acro{awg}[AWG]{Additive White Gaussian}
\acro{rhs}[r.h.s.]{right hand side}
\acro{lhs}[l.h.s.]{left hand side}
\acro{wrt}[w.r.t.]{with respect to}
\acro{lln}[LLN]{Law of Large Numbers}
\acro{clt}[CLT]{Central Limit Theorem}
\acro{sinr}[SINR]{Signal to Interference and Noise Ratio}
\end{acronym}
%
\title{RSB Decoupling Property of MAP Estimators\vspace{-2mm}}

\author{\IEEEauthorblockN{Ali Bereyhi\IEEEauthorrefmark{1}, Ralf M\"uller\IEEEauthorrefmark{1}, Hermann Schulz-Baldes\IEEEauthorrefmark{2}}
\IEEEauthorblockA{\IEEEauthorrefmark{1}Institute for Digital Communications (IDC), \IEEEauthorrefmark{2}Department of Mathematics,\\ Friedrich Alexander Universit\"at (FAU), Erlangen, Germany\\ ali.bereyhi@fau.de, ralf.r.mueller@fau.de, schuba@mi.uni-erlangen.de \vspace{-4mm}
\thanks{This work was supported by the German Research Foundation, Deutsche Forschungsgemeinschaft (DFG), under Grant No. MU 3735/2-1.}
}}

\IEEEoverridecommandlockouts
%
\maketitle

\tikzstyle{block} = [draw,  rectangle, minimum height=2.5em, minimum width=5em]
\tikzstyle{margin} = [draw, dotted, rectangle, minimum height=2.1em, minimum width=2em]
\tikzstyle{sum} = [draw, circle, node distance=1cm, inner sep=0pt]
\tikzstyle{input} = [coordinate]
\tikzstyle{output} = [coordinate]
\tikzstyle{pinstyle} = [pin edge={to-,thick,black}]
\begin{abstract}
The large-system decoupling property of a MAP estimator is studied when it estimates the i.i.d. vector $\bx$ from the observation $\by=\mA \bx+\bz$ with $\mA$ being chosen from a wide range of matrix ensembles, and the noise vector $\bz$ being i.i.d. and Gaussian. Using the replica method, we show that the marginal joint distribution of any two corresponding input and output symbols converges to a deterministic distribution which describes the input-output distribution of a single user system followed by a MAP estimator. Under the $\boldsymbol{b}$RSB assumption, the single user system is a scalar channel with additive noise where the noise term is given by the sum of an independent Gaussian random variable and $\boldsymbol{b}$ correlated interference terms. As the $\boldsymbol{b}$RSB assumption reduces to RS, the interference terms vanish which results in the formerly studied RS decoupling principle.
\end{abstract}
\section{Introduction}
A linear vector system with \ac{awgn} is described by\vspace{-1mm}
\begin{align}
\by=\mA \bx + \bz \label{eq:1}
\end{align}
where the \ac{iid} source vector $\bx_{n \times 1}$, taken from support $\maX^n$, is measured by the random system matrix $\mA_{k \times n}$ and corrupted by an \ac{iid} Gaussian noise vector $\bz_{k \times 1}$. The observation vector $\by$ is given to the vector estimator $\bgg(\cdot)$ which maps the $k$-dimensional vector $\by$ to an $n$-dimensional vector $\bhx_{n \times 1} \in \maX^n$. The entries of $\bhx$ are in general correlated due to the coupling imposed by $\mA$ and $\bgg(\cdot)$. Considering the entries $x_j$ and $\hx_j$, $1\leq j \leq n$, the marginal joint distribution of $(\hx_j,x_j)$ in the large-system limit, i.e. $k,n \uparrow \infty$, is of interest. To clarify the point, consider the linear estimation, i.e. $\bgg(\by)=\mG^{\trp} \by$ for some $\mG_{k \times n}$, and denote $\mA=[\ba_1 \cdots \ba_n]$ and $\mG=[\bgg_1 \cdots \bgg_n]$ with $\ba_i$ and $\bgg_i$ being $k \times 1$ vectors for $i \in \{ 1, \ldots, n\}$. Thus,
\begin{align}
\hx_j &= \left( \bgg_j^\trp \ba_j \right) x_j +  \sum_{\substack{i=1, i\neq j}}^n \left( \bgg_j^\trp \ba_i \right) x_i + \bgg_j^\trp \bz. \label{eq:2}
\end{align}
One considers the \ac{rhs} of \eqref{eq:2} as the linear estimation of a single user system with additive impairment in which the impairment term is not necessarily Gaussian and the system is indexed by $j$. For some families of $\mA$ and $\mG$, it is shown that the index dependency of these systems vanishes and the impairment term converges to a Gaussian noise term with modified power level when the system dimensions tend to infinity, e.g. \cite{guo2002asymptotic}. Thus, one can assume the linear vector estimator in the large-system limit to decouple into a bank of single user linear estimators operating over $n$ parallel scalar systems with additive Gaussian noise terms. This decoupling property of the linear estimators is rigorously justified invoking the central limit theorem and the properties of large random matrices. For nonlinear forms of $\bgg(\cdot)$, however, the analysis faces difficulties, since the output entries do not linearly decouple. Tanaka noted the similarity between the asymptotic analysis of spin glasses \cite{edwards1975theory} and vector estimators and showed that the performance of a vector estimator in the large-system limit can be represented as the macroscopic parameter of a spin glass \cite{tanaka2002statistical}. Consequently, a class of generally nonlinear estimators was analyzed using the nonrigorous replica method developed in statistical mechanics. Inspired by \cite{tanaka2002statistical}, several works employed the replica method to study the performance of nonlinear estimators in the asymptotic regime considering different classes of estimators, system matrices, and performance measures, e.g. \cite{muller2004capacity}. Having the decoupling property of the linear estimators in mind, it was conjectured that this property holds for nonlinear estimators as well. In \cite{guo2005randomly}, Guo and Verd\'u justified this conjecture for the postulated \ac{mmse} estimator \vspace{-.4mm}
\begin{align}
\bgg(\by)=\mathsf{E} \{\bx|\by,\mA\} \label{eq:3}
\end{align}
where $\mA$ is considered to be \ac{iid}, and the expectation is taken over $\bx$ due to some postulated posterior distribution $\mathrm{q}_{\bx|\by,\mA}$. For this setup, the authors showed the \ac{rs} decoupling principle which says that under the \ac{rs} assumption the marginal joint distribution of $(x_j,\hx_j)$ converges to the input-output joint distribution of a scalar channel with additive Gaussian noise followed by a single user \ac{mmse} estimator. The \ac{rs} decoupling principle was further extended to the case with a postulated \ac{map} estimator in \cite{rangan2012asymptotic} where Rangan et al. studied
\begin{align}
\bgg(\by)= \arg \min_\bv \ \left[ \frac{1}{2\lambda} \norm{\by-\mA \bv}^2 + u(\bv) \right] \label{eq:4}
\end{align}
for some ``utility function'' $u(\cdot): \real^n \rightarrow \real^+$ and non-negative real ``estimation parameter'' $\lambda$. Except for the cases with an \ac{iid} system matrix, the decoupling property of nonlinear estimators for a larger class of matrix ensembles has not yet been addressed precisely. In \cite{tulino2013support}, the authors investigated this issue partially by studying the support recovery of sparse Gaussian sources. They considered the case of a source vector which is first randomly measured by a squared matrix, and then, the measurements are sparsely sampled by an \ac{iid} binary vector. Employing a \ac{map} estimator for recovering, the \ac{rs} decoupling principle was justified for the case in which the measuring matrix belongs to a large set of matrix ensembles. Although the class of matrices is broadened in \cite{tulino2013support}, the result cannot be considered as a complete generalization of \cite{guo2005randomly} and \cite{rangan2012asymptotic}, since it is restricted to cases with a sparse Gaussian source and $kn^{-1} \leq 1$. Another issue not investigated in the literature is the marginal joint distribution under the \ac{rsb} assumption. In fact, the previous studies investigated the decoupling principle considering the \ac{rs} ansatz; however, despite the \ac{rs} validity in some particular cases, there are still several cases requiring further \ac{rsb} investigations, e.g. \cite{kabashima2009typical,zaidel2012vector}. 

In this paper, we address both the issues and broaden the scope of the decoupling principle stated in \cite{rangan2012asymptotic} to both a larger set of matrix ensembles, and the \ac{rsb} ans\"atze.
More precisely, we justify the decoupling property of the postulated \ac{map} estimator when
\begin{enumerate}
\item $\mA$ is chosen from a large family of random matrices,
\item $k n^{-1}$ takes any non-negative real number, and
\item the \ac{rs} and \ac{rsb} ans\"atze are considered.
\end{enumerate}
For this setup, we show that under all replica ans\"atze, the joint distribution of $(x_j,\hx_j)$ in the large-system limit converges to the input-output distribution of a scalar system in which the source symbol is corrupted by effective noise and estimated by a single user \ac{map} estimator. We determine the effective noise term and estimation parameter under the \ac{rsb} assumption with $b$ steps of breaking ($b$\ac{rsb}), and show that the noise term under this assumption is given by the sum of an independent Gaussian random variable and $b$ correlated terms. By reducing the assumption to \ac{rs}, the correlated terms vanish, and the noise term becomes Gaussian. Thus, one can consider the decoupling principle of \cite{rangan2012asymptotic} to be a special case of the more general decoupling principle illustrated here.

\textbf{Notation:} We represent vectors, scalars and matrices with bold lower case, non-bold lower case, and bold upper case letters, respectively. The set of real numbers is denoted by $\real$, and $\mA^{\trp}$ and $\mA^{\mathsf{H}}$ indicate the transposed and Hermitian of $\mA$. $\mI_m$ is the $m\times m$ identity matrix, $\mone_m$ is the matrix with all entries equal to one, and $\otimes$ denotes the Kronecker product. For a random variable $x$, $\mathrm{p}_x$ represents either the \ac{pmf} or \ac{pdf}, and $\mathrm{F}_x$ represents the \ac{cdf}. We denote the expectation over $x$ by $\mathsf{E}_x$, and an expectation over all random variables involved in a given expression by $\mathsf{E}$. For sake of compactness, the set of integers $\{1, \ldots,n \}$ is denoted by $[1:n]$, the zero-mean and unit-variance Gaussian \ac{pdf} by $\pi(\cdot)$, and 
\begin{align}
\int \md t \coloneqq \int \pi(t) \dif t.
\end{align}
Whenever needed, we consider the entries of $\bx$ to be discrete random variables; the results of this paper, however, are in full generality and directly extend to continuous distributions.


%
\section{Problem Formulation}
\label{sec:sys_setup}
Let the system in \eqref{eq:1} satisfy the following constraints.
\begin{enumerate}[label=(\alph*)]
\item The number of observations $k$ is a deterministic sequence of $n$ such that 
\begin{align}
\lim_{n \uparrow \infty} \frac{k}{n}=\frac{1}{\rate} < \infty.
\end{align}
\item $\bx_{n \times 1}$ is an \ac{iid} random vector with each element being distributed due to $\mathrm{p}_x$ over $\maX$ in which $\maX \subseteq \real$.
\item $\mA_{k \times n}$ is randomly generated over $\mathbbmss{A}^{k \times n} \subseteq \real^{k \times n}$, such that $\mJ=\mA^{\trp} \mA$ has the eigendecomposition 
\begin{align}
\mJ= \mU \mD \mU^{\trp} \label{eq:7}
\end{align}
where $\mU$ is an orthogonal Haar distributed matrix and $\mD$ is a diagonal matrix with the empirical eigenvalue distribution (density of states) converging as $n \uparrow \infty$ to a deterministic distribution $\mathrm{F}_{\mJ}$.
\item $\bz_{k \times 1}$ is a real \ac{iid} zero-mean Gaussian random vector with variance $\lambda_0$, i.e., $\bz \sim \man(\boldsymbol{0},\lambda_0 \mI_{k})$.
\item $\bx$, $\mA$, and $\bz$ are independent.
\end{enumerate}
In order to estimate the source vector, the postulated \ac{map} estimator as defined in \eqref{eq:4} is employed. The estimators postulates a non-negative estimation parameter $\lambda$ and a non-negative utility function $u(\cdot)$ which decouples, i.e., $u(\bx)=\sum_{i=1}^n u(x_i)$.

Defining the estimated vector $\bhx \coloneqq \bgg(\by)$, the conditional distribution of $\hx_j$ given $x_j$ for some $j \in [1:n]$ is denoted by $\prm_{\hx|x}^{j(n)}$. Thus, the marginal joint distribution of $x_j$ and $\hx_j$ at the mass point $(\hv,v)$ is written as
\begin{align}
\prm_{\hx_j,x_j}(\hv,v)=\prm_x(v) \prm_{\hx|x}^{j(n)}(\hv|v). \label{eq:8}
\end{align}
Considering the large-system limit, we define the asymptotic conditional distribution of $\hx_j$ given $x_j$ at $(\hv,v)$ as 
\begin{align}
\prm_{\hx|x}^{j} (\hv|v) \coloneqq \lim_{n \uparrow \infty} \prm_{\hx|x}^{j(n)} (\hv|v). \label{eq:9}
\end{align}
We also suppose the self averaging assumption which says
\begin{enumerate}[label=(\alph*)]
\addtocounter{enumi}{5}
\item Given $\mA$ of the form \eqref{eq:7} with $\mathrm{F}_{\mJ}$, the limit in \eqref{eq:9} exists and is almost surely constant in realizations of $\mA$.
\end{enumerate}
\section{General Decoupling Principle}
\label{sec:decoupling}
The main contribution of this study is to extend the scope of the decoupling principle. To illustrate the result, consider the following single user system: the input $x$ is passed through the channel $y=x+z$ where $z \sim \prm_{z|x}$ for the given input $x$. The observation $y$ is then given to a single user \ac{map} estimator with the same utility function as for the vector estimator defined in Section \ref{sec:sys_setup}, i.e. $u(\cdot)$, and an estimation parameter denoted by $\lams$. Indicating the conditional distribution of the estimator's output $\hx$ for the given input $x$ by $\prm_{\hx|x}$, our general decoupling principle says that under a set of assumptions
\begin{enumerate}[label=(\alph*)]
\item the asymptotic conditional distribution $\prm^{j}_{\hx|x}$ is independent of the index $j$, and we have $\prm_{\hx|x}^{j}= \prm_{\hx|x}$.
\item $\prm_{z|x}$ and $\lams$ are determined in terms of $\lambda$, $\lambda_0$ and the statistics of $\bx$ and $\mA$.
\end{enumerate}
The set of assumptions which yields the validity of the above statements are enforced within the large-system analysis. In the following, we briefly illustrate our approach and determine the parameters of the single user system.
\section{Derivation of General Decoupling Principle}
\label{sec:derivation}
\label{subsec:derivation}
Before illustrating our derivation approach, let us define the $\mathrm{R}$-transform. For a random variable $t$, the Stieltjes transform over the upper half complex plane is defined as $\mathrm{G}_t(s)= \E [t-s]^{-1}$. Denoting the inverse \ac{wrt} composition with $\mathrm{G}_t^{-1}(\cdot)$, the $\mathrm{R}$-transform is $\mathrm{R}_t(\omega)= \mathrm{G}_t^{-1}(\omega) - \omega^{-1}$ such that $\lim_{\omega \downarrow 0} \mathrm{R}_{t}(\omega) = \E t$. The definition also extends to matrix arguments. Assuming a matrix $\mM_{n \times n}$ to have the eigendecomposition $\mM=\mU \ \mathrm{diag}[\lambda_1, \ldots, \lambda_n] \ \mU^{-1}$, $\mathrm{R}_t(\mM)$ is then defined as $\mathrm{R}_t(\mM)=\mU \ \mathrm{diag}[\mathrm{R}_t(\lambda_1), \ldots, \mathrm{R}_t(\lambda_n)] \ \mU^{-1}$.

The derivation of the general decoupling principle is based on the moment method. To clarify the approach, consider the non-negative integers $k$ and $\ell$, and define the joint moment $\Mrm_{k,\ell}^{j(n)}=\E \hx_j^k x_j^\ell$, for $j \in [1:n]$. After evaluating the limit of $\Mrm_{k,\ell}^{j(n)}$ as $n \uparrow \infty$, we show that for all $k$ and $\ell$ the asymptotic joint moment is equivalent to the corresponding joint moment of the single user system. Consequently, using the uniqueness of the mapping from the set of integer moments' sequences to the set of measures, under a set of conditions investigated in the classical moment problem \cite{akhiezer1965classical}, we conclude that both couples $(\hx_j , x_j)$ and $(\hx , x)$ have a same distribution. We start with evaluating the limit of $\Mrm_{k,\ell}^{j(n)}$. The evaluation is based on the nonrigorous method of replicas developed in the theory of spin glasses \cite{edwards1975theory}, and accepted as a mathematical tool in information theory. To do so, define the ``weighted average joint moment'' over the index set $\maw \subset [1:n]$ as
\begin{align}
\moment_{k,\ell}^{\maw(n)} (\bhx;\bx) \coloneqq \E \frac{1}{\abs{\maw}} \sum_{w \in \maw} \hx_w^k x_w^\ell. \label{eq:12}
\end{align}
Setting $\maw=[j:j+ n \eta]$ for some $\eta \in (0,1]$, the asymptotic joint moment of $j$th entry can be written as
\begin{align}
\Mrm_{k,\ell}^{j}\coloneqq\lim_{n \uparrow \infty} \Mrm_{k,\ell}^{j(n)}= \lim_{n \uparrow \infty}\lim_{\eta \downarrow 0} \moment_{k,\ell}^{\maw(n)} (\bhx;\bx). \label{eq:13}
\end{align}
Thus, the evaluation of the asymptotic moment reduces to taking the limits in the \ac{rhs} of \eqref{eq:13} which needs the weighted average joint moment in \eqref{eq:12} to be explicitly calculated for an arbitrary integer $n$. Alternatively, we can define the function
\begin{align}
\maz(\beta,h) = \sum_{\bv}  e^{ -\beta \left[ \frac{1}{2\lambda} \norm{\by-\mA \bv}^2 + u(\bv) \right]+h n\moment_{k,\ell}^{\maw(n)} (\bv;\bx)}. \label{eq:14}
\end{align}
with $\bv \in \maX^n$. Noting that $\bhx=\bgg(\by)$ with $\bgg(\cdot)$ defined in \eqref{eq:4},
\begin{align}
\moment_{k,\ell}^{\maw(n)} (\bhx;\bx)=\lim_{\beta \uparrow \infty} \lim_{h \downarrow 0} \frac{1}{n}\frac{\partial}{\partial h} \ \E \log \maz(\beta,h). \label{eq:15}
\end{align}
The logarithmic expectation in the \ac{rhs} of \eqref{eq:15} is not a trivial task to do, and therefore, one bypasses the direct evaluation using the Riesz equality which for any random variable $t$ states $\E \log t = \lim_{m \downarrow 0} m^{-1}\log \E t^m$. Thus, regarding \eqref{eq:13} and \eqref{eq:15}
\begin{align}
\Mrm_{k,\ell}^{j}= \lim_{n \uparrow \infty}\lim_{\eta \downarrow 0} \lim_{\beta \uparrow \infty} \lim_{h \downarrow 0} \lim_{m \downarrow 0} \frac{1}{n} \frac{\partial}{\partial h} \ \frac{\log \E \left[\maz(\beta,h)\right]^m}{m} \label{eq:16}
\end{align}
for $\maw=[j:j+ n \eta]$. In \eqref{eq:16}, we face two major difficulties:
\begin{inparaenum}
\item evaluating the real moments i.e., $\E\left[\maz(\beta,h)\right]^m$, and
\item taking the limits in the order stated.
\end{inparaenum}
Basic analytical methods fail to address these challenges properly, and therefore, we invoke the nonrigorous method of replicas. The replica method suggests to evaluate the moment for an arbitrary integer $m$ as an analytic function in $m$; then, assume that
\begin{inparaenum}
\item the ``replica continuity'' holds which means that the function analytically continues from the set of integers to the real axis (or at least a vicinity of zero), and
\item the limits are exchangeable.
\end{inparaenum}
Following the above prescription, we consider the first assumption and find $\E\left[\maz(\beta,h)\right]^m$ which for an integer $m$ reduces to
\begin{align}
\hspace{-1mm} \mathsf{E}\prod_{a=1}^m \sum_{\bv_a} e^{-\beta \left[ \frac{1}{2\lambda} \norm{\mA (\bx-\bv_a) + \bz}^2 + u(\bv_a) \right]+h n \moment_{k,\ell}^{\maw(n)} (\bv_a;\bx)}. \label{eq:17}
\end{align}
In order to evaluate \eqref{eq:17}, one can initially take the expectation over $\bz$ and $\mA$. Due to the lack of space, we leave the details for the extended version of the manuscript; however, we briefly explain the strategy. After taking the expectations, and defining the ${m \times m}$ ``replica correlation matrix'' $\mQ$ such that $[\mQ]_{ab} = n^{-1} (\bx-\bv_a)^\trp (\bx-\bv_b)$, \eqref{eq:17} is given in terms of $\mQ$ as
\vspace{-1mm}
\begin{align}
\E \left[ \maz(\beta,h) \right]^m = \mathsf{E}_{\bx} \int e^{-n\mg (\mT\mQ)} e^{n \mai(\mQ)} \dif \mQ \label{eq:19}
\end{align}
with $\dif \mQ \coloneqq \prod_{a,b=1}^{m} \dif [\mQ]_{ab}$, $\mT\coloneqq \frac{1}{2 \lambda}\mI_m-\beta \frac{\lambda_0}{2\lambda^2} \mone_m$, and the integral being taken over $\real^{m^2}$. For a given $\bx$, $e^{n\mai(\mQ)}$ measures the probability weight of the set of replicas, $\{ \bv_a \}_{a=1}^m$, in which the correlation matrix is $\mQ$; moreover, $\mg(\cdot)$ is defined as
\vspace{-1mm}
\begin{align}
\mg(\mM) =  \int_{0}^{{\beta}} \Tr \{\mM \mathrm{R}_{\mJ}(-2\omega\mM)\} \dif \omega  + \epsilon_n \label{eq:20}
\end{align}
where $\Tr\{ \cdot \}$ denotes the trace, $\mathrm{R}_{\mJ}(\cdot)$ is the $\mathrm{R}$-transform \ac{wrt} $\mathrm{F}_{\mJ}$, and $\epsilon_n$ tends to zero as $n \uparrow \infty$.
Here, one can employ the Laplace method of integration and replace the \ac{rhs} of \eqref{eq:19} in the large-system limit with the integrand at its saddle point multiplied by some bounded coefficient $\mathsf{K}_n$; thus, as $n \uparrow \infty$\vspace{-.65mm}
\begin{align}
\E \left[ \maz(\beta,h) \right]^m \doteq \mathsf{K}_n e^{-n \left[ \mg (\mT\tilde{\mQ})-\mai (\tilde{\mQ}) \right]} \label{eq:23}
\end{align}
at the saddle point $\tilde{\mQ}$.
Substituting \eqref{eq:23} in \eqref{eq:16}, we have\vspace{-.5mm}
\begin{align}
\Mrm_{k,\ell}^{j}=\lim_{\eta \downarrow 0}  \lim_{\beta \uparrow \infty} \lim_{m \downarrow 0}  \E  \frac{\sum_{\bvv} \moment_{k,\ell}^{[1:m]} (\bvv;\bxx) \mae(\tilde{\mQ},\bxx,\bvv;\beta)}{ \sum_{\bvv} \mae(\tilde{\mQ},\bxx,\bvv;\beta)} \label{eq:25}
\end{align}
where $\bxx_{m \times 1}=[x, \ldots, x]^{\trp}$ with $x \sim \prm_x$, $\bvv_{m \times 1} \in \maX^m$ and
\begin{align}
\mae(\tilde{\mQ},\bxx,\bvv;\beta)= e^{- \beta (\bxx-\bvv)^{\trp} \mT \mathrm{R}_{\mJ}(-2\beta\mT \tilde{\mQ}) (\bxx - \bvv) - \beta u(\bvv)}. \label{eq:22}
\end{align}
Here, one needs to find the saddle point which is not a feasible task in general. The strategy for pursuing the analysis is to restrict the saddle point, i.e. $\tilde{\mQ}$, to be of a special form, and find the solution within the restricted set of matrices. This is where an additional assumption such as the \ac{rs} or \ac{rsb} assumption arises. It is clear that these restrictions do not lead us to the correct solution in general, and therefore, one needs to widen the set of replica correlation matrices to find a more accurate solution. In the sequel, we consider different structures on the correlation matrix and find the replica ansatz under those assumptions. However, considering \eqref{eq:25}, it is observed that ``regardless of the structure'' on $\tilde{\mQ}$, the joint moment $\moment_{k,\ell}^j$ is independent of the index $j$ even before taking the limit $\eta \downarrow 0$. Therefore, employing the moment method, we conclude
\begin{proposition}
\label{prop:1}
Let the vector system satisfy the constraints in Section \ref{sec:sys_setup}; moreover, assume the replica continuity to hold, and the limits in \eqref{eq:16} to exist and exchange. Then $\prm_{\hx|x}^j$, as defined in \eqref{eq:9}, does not depend on the index $j$.
\end{proposition}
\vspace{-2mm}
Proposition \ref{prop:1} states a more general form of the decoupling principle studied in previous works. In fact, the only assumptions which need to be satisfied are the replica continuity and the exchange of limits; and, no structure of the correlation matrix is imposed. However, the decoupled scalar system does depend on the structure imposed on $\tilde{\mQ}$. To find the decoupled single user system, we start with the most primary structure which is imposed by considering the \ac{rs} assumption.\\
\textbf{\ac{rs} Assumption:} Here, we restrict the search to the set of parameterized matrices which are of the form
\begin{align}
\tilde{\mQ}=q \mone_m + \frac{\chi}{\beta} \mI_m.\vspace{-1mm} \label{eq:26}
\end{align}
for some $\chi , q \in \real^+$. Substituting in \eqref{eq:25}, we have\vspace{-1mm}
\begin{align}
\Mrm_{k,\ell}^j= \E \int \mathrm{g}^k x^{\ell} \md z \label{eq:27}
\end{align}
where $\mathrm{g}\coloneqq\mathrm{g}_{\mathsf{map}}[(y);\lams,u]$ with $y=x+\sqrt{\lams_0} z$ and\vspace{-1mm}
\begin{align}
\mathrm{g}_{\mathsf{map}}[(y);\lams,u]=\arg \min_{v} \left[ \frac{1}{2\lams} (y -v)^2+u(v) \right]. \label{eq:28}
\end{align}
Moreover, $\lams_0$ and $\lams$ are defined as
\begin{subequations}
\begin{align}
\lams_0&=\left[ \mathrm{R}_{\mJ}(-\frac{\chi}{\lambda}) \right]^{-2} \frac{\partial}{\partial \chi} \left\lbrace \left[ \lambda_0 \chi - \lambda q \right] \mathrm{R}_{\mJ}(-\frac{\chi}{\lambda}) \right\rbrace \label{eq:29a} \\
\lams&= \left[ \mathrm{R}_{\mJ}(-\frac{\chi}{\lambda}) \right]^{-1} \lambda \label{eq:29b}
\end{align}
\end{subequations}
where $q = \E \int [\mathrm{g}-x]^2 \md z$, and $\chi$ satisfies\vspace{-1mm}
\begin{align}
\sqrt{\lams_0} \chi &= \lams \E \int [\mathrm{g}-x] z \md z. \vspace{-2mm} \label{eq:30a}
\end{align}
\eqref{eq:28} describes a single user \ac{map} estimator with the postulated utility function $u(\cdot)$ and estimation parameter $\lams$. Thus,
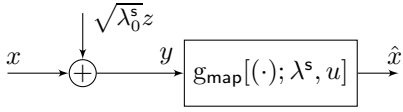
\begin{figure}
\begin{center}
\begin{tikzpicture}[auto, node distance=2.5cm,>=latex']
    \node [input, name=input] {};
    \node [sum, right of=input] (sum) {$+$};
    \node [input, above of=sum, node distance=.8cm] (noise) {};
    \node [block, right of=sum] (estimator) { $\mathrm{g}_{\mathsf{map}}[(\cdot);\lams,u]$ };
    \node [output, right of=estimator, node distance=1.7cm] (output) {};

    \draw [draw,->] (input) -- node[pos=.1] {$x$} (sum) ;
    \draw [->] (sum) -- node[pos=.8] {$y$} (estimator);
    \draw [->] (estimator) -- node[pos=.9] [name=x_h] {$\hx$}(output);
	\draw [draw,->] (noise) -- node[pos=.1] {\small{$\sqrt{\lams_0}$}$z$}(sum);
\end{tikzpicture}
\end{center}
	\caption{The decoupled scalar system under the \ac{rs} ansatz.}
	\label{fig:1}
\end{figure}
\vspace{-2mm}
\begin{proposition}
\label{prop:2}
Let the assumptions in Proposition \ref{prop:1}, as well as the \ac{rs} assumption hold, and consider the single user system in Figure \ref{fig:1} with $\lams_0$ and $\lams$ as defined in \eqref{eq:29a} and \eqref{eq:29b}. Then for $j \in [1:n]$, $\prm_{\hx|x}^j$ as defined in \eqref{eq:9} describes the conditional distribution of $\hx$ given $x$ in Figure \ref{fig:1} where $\mathrm{g}_{\mathsf{map}}[(\cdot);\lams,u]$ is a single user \ac{map} defined in \eqref{eq:28}, and $\prm_{z|x}(z|x)=\pi(z)$.\vspace{-2mm}
\end{proposition}
The results in the literature have always considered the \ac{rs} ansatz, and can be recovered as special cases of Proposition \ref{prop:2}. E.g., results of \cite{rangan2012asymptotic} are derived by setting $\mathrm{R}_{\mJ}(\omega)=(1-\rate \omega)^{-1}$. The \ac{rs} ansatz, however, does not provide a valid solution, in general. Parisi in \cite{parisi1980sequence} introduced the \ac{rsb} scheme which widens the restricted set of saddle point matrices recursively. To illustrate the \ac{rsb} scheme, let $\mQ^{\mathsf{b}}$ be a basic structure for the replica correlation matrix; moreover, assume $m$ to be a multiple of an integer $\xi$. Then, the correlation matrix can be grouped as a $\xi \times \xi$ matrix of blocks with each block being an $\frac{m}{\xi} \times \frac{m}{\xi}$ matrix. In this case, a new structure for the correlation matrix is obtained by setting the diagonal blocks to be $\mQ^{\mathsf{b}}$ and the off-diagonal blocks to be $\kappa \mone_{\frac{m}{\xi}}$ for some $\kappa$. In fact, the new set of correlation matrices is constructed by imposing the \ac{rs} structure block-wisely using $\mQ^{\mathsf{b}}$ and $\kappa \mone_{\frac{m}{\xi}}$ as basic blocks. The \ac{rsb} scheme can be recursively iterated: one can set the basic structure  $\mQ^{\mathsf{b}}$ and find the new structure $\mQ^{\mathsf{b}}_1$; then, by taking $\mQ^{\mathsf{b}}_1$ as the new basic structure, a wider set of correlation matrices is found. Parisi considers $\mQ^{\mathsf{b}}$ to have the \ac{rs} structure.\\
\textbf{1\ac{rsb} Assumption:} Using the \ac{rsb} scheme with one step of iteration, the structure of the correlation matrix is found as\vspace{-1mm}
\begin{align}
\tilde{\mQ}= q \mone_m + p \mI_{\frac{m \beta}{\mu}} \otimes \mone_{\frac{\mu}{\beta}} + \frac{\chi}{\beta} \mI_m \vspace{-1mm} \label{eq:31}
\end{align}
for some $\chi , p, q, \mu \in \real^+$. Therefore, the joint moment reads\vspace{-1mm}
\begin{align}
\Mrm_{k,\ell}^j= \E \int \mathrm{g}^k x^{\ell} \dif \mathrm{F}(z_1) \md z_0\vspace{-1mm} \label{eq:32}
\end{align}
where $\mathrm{g}\coloneqq\mathrm{g}_{\mathsf{map}}[(y);\lams,u]$ with $y=x+\sqrt{\lams_0} z_0+\sqrt{\lams_1} z_1$, and $\dif \mathrm{F}(z_1)=\Lambda \md z_1$ with $\Lambda=[\int \tilde{\Lambda} \md z_1]^{-1}\tilde{\Lambda}$ and
\begin{align}
\tilde{\Lambda}=e^{-\mu \left\lbrace \tfrac{1}{2 \lams} \left[ (y-\mathrm{g})^2-(y-x)^2 \right]+ u(\mathrm{g}) \right\rbrace}. \label{eq:34}
\end{align}
Moreover, by denoting $\varrho \coloneqq \chi+\mu p$, we have
\begin{subequations}
\begin{align}
\lams_0 &= \left[ \mathrm{R}_{\mJ}(-\frac{\chi}{\lambda}) \right]^{-2} \frac{\partial}{\partial \varrho} \left\lbrace \left[\lambda_0\varrho-\lambda q+\lambda p \right] \mathrm{R}_{\mJ}(- \frac{\varrho}{\lambda})\right\rbrace, \label{eq:35a} \\
\lams_1 &=\left[ \mathrm{R}_{\mJ}(-\frac{\chi}{\lambda}) \right]^{-2} \left[ \mathrm{R}_{\mJ}(- \frac{\chi}{\lambda}) - \mathrm{R}_{\mJ}(- \frac{\varrho}{\lambda}) \right] \lambda \mu^{-1}, \label{eq:35b} \\
\lams&= \left[ \mathrm{R}_{\mJ}(-\frac{\chi}{\lambda}) \right]^{-1} \lambda. \label{eq:35c}
\end{align}
\end{subequations}
Here, $q =  \E \int [\mathrm{g}-x]^2 \dif \mathrm{F}(z_1) \md z_0$, and $\chi$ and $p$ satisfy\vspace{-1mm}
\begin{subequations}
\begin{align}
\chi+\mu p&= \frac{\lams}{\sqrt{\lams_0}} \E \int [\mathrm{g}-x] z_0 \dif \mathrm{F}(z_1) \md z_0 \label{eq:36a} \\
\chi+\mu q&= \frac{\lams}{\sqrt{\lams_1}} \E \int [\mathrm{g}-x] z_1 \dif \mathrm{F}(z_1) \md z_0 \label{eq:36b}
\end{align}
\end{subequations}
for some $\mu$ being a solution to the fixed point equation
\begin{align}
\frac{\mu}{2\lams}&\left[\mu\frac{\lams_1}{\lams}q-\mu\frac{\lams_1}{\lams}p+p\right]-\frac{1}{2\lambda}\int_\chi^\varrho \mathrm{R}_{\mJ}(-\frac{\omega}{\lambda}) \dif \omega = \qquad \nonumber \\ 
&\qquad = \ \mathrm{I}(z_1;x,z_0) + \mathsf{D}_{\mathsf{KL}}(\prm_{z_1} \Vert \pi) \label{eq:37}
\end{align}
where $\mathrm{I}(\cdot;\cdot)$ and $\mathsf{D}_{\mathsf{KL}}(\cdot \Vert \cdot)$ indicate the mutual information and the so-called ``Kullback-Leibler'' distance respectively, and the random variables $(x,z_0,z_1) \sim \prm_{x}(x)\pi(z_0)\left[ \Lambda \pi(z_1) \right]$. Thus, one can conclude the following proposition.\vspace{-2mm}
\begin{figure}
\begin{center}
\begin{tikzpicture}[auto, node distance=2.5cm,>=latex']
    \node [input, name=input] {};
    \node [sum, right of=input] (sum0) {$+$};
    \node [sum, right of=sum0, node distance=1.4cm] (sum1) {$+$};
    \node [margin, right of=sum0, node distance=1.4cm,minimum width=2.8em] (margin) {};
    \node [input, above of=sum0, node distance=.8cm] (noise0) {};
    \node [input, above of=sum1, node distance=.8cm] (noise1) {};    
    \node [block, right of=sum1] (estimator) { $\mathrm{g}_{\mathsf{map}}[(\cdot);\lams,u]$ };
    \node [output, right of=estimator, node distance=1.7cm] (output) {};

    \draw [draw,->] (input) -- node[pos=.1] {$x$} (sum0) ;
    \draw [->] (sum0) -- node { } (sum1);
    \draw [->] (sum1) -- node[pos=.8] {$y$} (estimator);
    \draw [->] (estimator) -- node[pos=.9] [name=x_h] {$\hx$}(output);
	\draw [draw,->] (noise0) -- node[pos=.1] {\small $\sqrt{\lams_0}z_0$}(sum0);
	\draw [draw,->] (noise1) -- node[pos=.1] {\small $\sqrt{\lams_1}z_1$}(sum1);	
\end{tikzpicture}
\end{center}
	\caption{The decoupled scalar system under the \ac{1rsb} ansatz.}
	\label{fig:2}
\end{figure}
\begin{proposition}
\label{prop:3}
Let the assumptions in Proposition \ref{prop:1}, as well as the 1\ac{rsb} assumption hold, and consider the single user system in Figure \ref{fig:2} with $\lams_0$, $\lams_1$ and $\lams$ as defined in \eqref{eq:35a}-\eqref{eq:35c}. Then, for $j \in [1:n]$, $\prm_{\hx|x}^j$ in \eqref{eq:9} is the conditional distribution of $\hx$ given $x$ in Figure \ref{fig:2} where $\mathrm{g}_{\mathsf{map}}[(\cdot);\lams,u]$ is a single user \ac{map} estimator defined in \eqref{eq:28}, $\prm_{z_0|x}(z_0|x)=\pi(z_0)$, and
\begin{align}
\prm_{z_1|x,z_0}(z_1|x,z_0)= \Lambda\pi(z_1)
\end{align}
with $\Lambda=[\int \tilde{\Lambda} \md z_1]^{-1}\tilde{\Lambda}$ and $\tilde{\Lambda}$ defined in \eqref{eq:34}. 
\end{proposition}
Here, the decoupled system differs from the system obtained under the \ac{rs} ansatz within one additive tap which is intuitively approximating the interference caused by the coupling. In fact, the \ac{rs} ansatz assumes the coupling caused by the system matrix and vector estimator to vanish as the system tends to its large limits; however, the 1\ac{rsb} solution takes the coupling into account and approximates it with one tap of interference. This approximation may become more accurate, if we let the correlation matrix to be chosen from a larger set of matrices.\\
\textbf{$\boldsymbol{b}$\ac{rsb} Assumption:} Iterating the \ac{rsb} scheme with $b$ steps,
\begin{align}
\tilde{\mQ}= q \mone_m + \sum_{\nu=1}^b p_\nu \mI_{\frac{m \beta}{\mu_\nu}} \otimes \mone_{\frac{\mu_\nu}{\beta}} + \frac{\chi}{\beta} \mI_m 
\end{align}
for some $\chi ,q, \{ p_\nu , \mu_\nu \}_{\nu=1}^b \in \real^+$. Thus, we have
\begin{align}
\Mrm_{k,\ell}^j= \E \int \mathrm{g}^k x^{\ell} \prod_{\nu=1}^{b} \dif \mathrm{F}(z_\nu) \md z_0  \label{eq:39}
\end{align}
where $\mathrm{g}\coloneqq\mathrm{g}_{\mathsf{map}}[(y);\lams,u]$ with $y=x+\sum_{\nu=0}^b \sqrt{\lams_\nu} z_\nu$, and $\dif \mathrm{F}(z_\nu)=\Lambda_\nu \md z_\nu$. For $\nu \in [1:b]$, $\Lambda_\nu$ is a function of $x$ and $\{z_\zeta \}_{\zeta=0}^\nu$. Due to the page limitations, we leave the expressions of $\lams$, $\{ \lams_\nu \}_{\nu=0}^b$ and $\{ \Lambda_\nu\}_{\nu=1}^b$ for the extended version.
\begin{proposition}
\label{prop:4}
Let the assumptions in Proposition \ref{prop:1}, and the $b$\ac{rsb} assumption hold; moreover, consider the single user system in Figure \ref{fig:3}. Then, for $j \in [1:n]$, $\prm_{\hx|x}^j$ as defined in \eqref{eq:9} describes the conditional distribution of $\hx$ given $x$ in Figure \ref{fig:3} where $\mathrm{g}_{\mathsf{map}}[(\cdot);\lams,u]$ is a single user \ac{map} estimator defined in \eqref{eq:28}, $\prm_{z_0|x}(z_0|x)=\pi(z_0)$, and
\begin{align}
\prm_{z_\nu|x,\{ z_{\zeta} \}_{\zeta=0}^{\nu-1}}(z_\nu|x,\{ z_{\zeta} \}_{\zeta=0}^{\nu-1})= \Lambda_\nu \pi(z_\nu)
\end{align}
for $\nu \in [1:b]$. The factor $\Lambda_\nu$ depends on $x$ and $\{z_\zeta \}_{\zeta=0}^\nu$, and the coefficients $\lams$ and $\{ \lams_\nu \}_{\nu=0}^b$ are coupled due to a set of fixed point equations and bounded as $b \uparrow \infty$.
\end{proposition}
\begin{figure}
\begin{center}
\begin{tikzpicture}[auto, node distance=2.5cm,>=latex']
    \node [input, name=input] {};
    \node [sum, right of=input] (sum0) {$+$};
    \node [sum, right of=sum0,  node distance=1.4cm] (sum1) {$+$};
    \node [margin, right of=sum1, node distance=.8cm,minimum width=2.6cm] (margin) {};
    \node [input, right of=sum1, node distance=.6cm] (int_node1) {};
    \node [input, right of=int_node1, node distance=.5cm] (int_node2) {};
    \node [sum, right of=sum1, node distance=1.6cm] (sumb) {$+$};
    \node [input, above of=sum0, node distance=.8cm] (noise0) {};
    \node [input, above of=sum1, node distance=.8cm] (noise1) {}; 
    \node [input, above of=sumb, node distance=.8cm] (noiseb) {}; 
    \node [block, right of=sumb] (estimator) { $\mathrm{g}_{\mathsf{map}}[(\cdot);\lams,u]$ };
    \node [output, right of=estimator, node distance=1.7cm] (output) {};

    \draw [draw,->] (input) -- node[pos=.1] {$x$} (sum0) ;
    \draw [->] (sum0) -- node { } (sum1);
    \draw [-,dotted] (int_node1) -- node { } (int_node2);    
    \draw [->] (sumb) -- node[pos=.8] {$y$} (estimator);
    \draw [->] (estimator) -- node[pos=.9] [name=x_h] {$\hx$}(output);
	\draw [draw,->] (noise0) -- node[pos=.1] {\small $\sqrt{\lams_0}z_0$}(sum0);
	\draw [draw,->] (noise1) -- node[pos=.1] {\small $\sqrt{\lams_1}z_1$}(sum1);	
	\draw [draw,->] (noiseb) -- node[pos=.1] {\small $\sqrt{\lams_b}z_b$}(sumb);	
\end{tikzpicture}
\end{center}
	\caption{The decoupled scalar system under the \ac{brsb} ansatz.}
	\label{fig:3}
\end{figure}

Considering the $b$\ac{rsb} ansatz, one concludes that the ansatz extends the decoupled system in Figure \ref{fig:2} by approximating the coupling interference with more taps. The approximation, however, stops to improve at some step $b^*$, if $\Lambda_\nu=1$ for any integer $\nu > b^*$. The extreme case is when for any $\nu \in [1:b]$ in the $b$\ac{rsb} ansatz $\Lambda_\nu=1$. Here, the random variables $\{z_\nu\}_{\nu=1}^b$ in Figure \ref{fig:3} become independent Gaussian, and therefore, the decoupled system reduces to Figure \ref{fig:1}. In fact in this case, the $b$\ac{rsb} solution, as well as any $\nu$\ac{rsb} ansatz with $\nu \in [1:b]$, reduces to the \ac{rs} ansatz. Thus, one can consider the decoupled system under the \ac{rs} ansatz to be a special case of the more general decoupled system given in Figure \ref{fig:3}.
\section{Conclusion}
\label{sec:conclusion} 
Decoupling seems to be a generic property of \ac{map} estimators, as Proposition \ref{prop:1} justifies it for any source distribution and a wide range of matrix ensembles. The validity of the result relies only on replica continuity; however, the equivalent single user system depends on the structure of the replica correlation matrix. Recent results in statistical mechanics have shown that failures in finding the exact solution via the replica method are mainly caused by the assumed structure, and not replica continuity. Inspired by the Sherrington-Kirkpatrick model of spin glasses, for which the $\infty$\ac{rsb} ansatz has been proved to be correct, one may consider Figure \ref{fig:3} to be the general decoupled system as $b \uparrow \infty$. However, in many cases an accurate approximation might be provided by a finite number of \ac{rsb} steps. An extreme case is the \ac{rs} ansatz where all the interference terms in the \ac{rsb} decoupled system become independent and Gaussian. Thus, one concludes that the previous results in the literature were both special and extreme cases of the \ac{rsb} decoupled system. The \ac{rsb} decoupled system raises several issues which require further investigations. For example, nothing is known about the distance between the conditional distributions of the interference terms and independent Gaussian distributions in probability space. The distance variation \ac{wrt} the number of interference taps can then describe the improvement caused by increasing the number of \ac{rsb} steps. \vspace{-1.1mm}

\bibliographystyle{IEEEtran}
\bibliography{ref}

\begin{thebibliography}{10}
\providecommand{\url}[1]{#1}
\csname url@samestyle\endcsname
\providecommand{\newblock}{\relax}
\providecommand{\bibinfo}[2]{#2}
\providecommand{\BIBentrySTDinterwordspacing}{\spaceskip=0pt\relax}
\providecommand{\BIBentryALTinterwordstretchfactor}{4}
\providecommand{\BIBentryALTinterwordspacing}{\spaceskip=\fontdimen2\font plus
\BIBentryALTinterwordstretchfactor\fontdimen3\font minus
  \fontdimen4\font\relax}
\providecommand{\BIBforeignlanguage}[2]{{%
\expandafter\ifx\csname l@#1\endcsname\relax
\typeout{** WARNING: IEEEtran.bst: No hyphenation pattern has been}%
\typeout{** loaded for the language `#1'. Using the pattern for}%
\typeout{** the default language instead.}%
\else
\language=\csname l@#1\endcsname
\fi
#2}}
\providecommand{\BIBdecl}{\relax}
\BIBdecl

\bibitem{guo2002asymptotic}
D.~Guo, S.~Verd{\'u}, and L.~K. Rasmussen, ``Asymptotic normality of linear
  multiuser receiver outputs,'' \emph{Information Theory, IEEE Transactions
  on}, vol.~48, no.~12, pp. 3080--3095, 2002.

\bibitem{edwards1975theory}
S.~F. Edwards and P.~W. Anderson, ``Theory of spin glasses,'' \emph{Journal of
  Physics F: Metal Physics}, vol.~5, no.~5, p. 965, 1975.

\bibitem{tanaka2002statistical}
T.~Tanaka, ``A statistical-mechanics approach to large-system analysis of CDMA
  multiuser detectors,'' \emph{Information Theory, IEEE Transactions on},
  vol.~48, no.~11, pp. 2888--2910, 2002.

\bibitem{muller2004capacity}
R.~R. M{\"u}ller and W.~H. Gerstacker, ``On the capacity loss due to separation
  of detection and decoding,'' \emph{Information Theory, IEEE Transactions on},
  vol.~50, no.~8, pp. 1769--1778, 2004.

\bibitem{guo2005randomly}
D.~Guo and S.~Verd{\'u}, ``Randomly spread cdma: Asymptotics via statistical
  physics,'' \emph{Information Theory, IEEE Transactions on}, vol.~51, no.~6,
  pp. 1983--2010, 2005.

\bibitem{rangan2012asymptotic}
S.~Rangan, A.~K. Fletcher, and V.~Goyal, ``Asymptotic analysis of MAP
  estimation via the replica method and applications to compressed sensing,''
  in \emph{IEEE Trans. on Inf. Theory}, 2012, pp. 1902--1923.

\bibitem{tulino2013support}
A.~M. Tulino, G.~Caire, S.~Verdu, and S.~Shamai, ``Support recovery with
  sparsely sampled free random matrices,'' \emph{Information Theory, IEEE
  Transactions on}, vol.~59, no.~7, pp. 4243--4271, 2013.

\bibitem{kabashima2009typical}
Y.~Kabashima, T.~Wadayama, and T.~Tanaka, ``A typical reconstruction limit for
  compressed sensing based on $\ell_p$-norm minimization,'' \emph{Journal of
  Statistical Mechanics: Theory and Experiment}, vol. 2009, no.~09, p. L09003,
  2009.

\bibitem{zaidel2012vector}
B.~M. Zaidel, R.~M\"uller, A.~L. Moustakas, and R.~de~Miguel, ``Vector
  precoding for gaussian MIMO broadcast channels: Impact of replica symmetry
  breaking,'' \emph{Information Theory, IEEE Transactions on}, vol.~58, no.~3,
  pp. 1413--1440, 2012.

\bibitem{akhiezer1965classical}
N.~I. Akhiezer, \emph{The classical moment problem: and some related questions
  in analysis}.\hskip 1em plus 0.5em minus 0.4em\relax Oliver \& Boyd, 1965,
  vol.~5.

\bibitem{parisi1980sequence}
G.~Parisi, ``A sequence of approximated solutions to the SK model for spin
  glasses,'' \emph{Journal of Physics A: Mathematical and General}, vol.~13,
  no.~4, p. L115, 1980.

\end{thebibliography}
%
%


\end{document}